\documentclass[aps, prb, twocolumn, showpacs, groupedaddress, showkeys]{revtex4}
\usepackage[dvips]{color}
\usepackage{xcolor}
\usepackage{amsfonts}
\usepackage{amsmath}
\usepackage{amssymb}
\usepackage{graphicx}
\usepackage{mathrsfs}
\usepackage{bbm}
\usepackage{bm}
\usepackage{multirow}
\usepackage[sort&compress]{natbib}
\usepackage{dcolumn}%

\begin{document}

\newcommand{\dd}{d}
\newcommand{\pd}{\partial}
\newcommand{\myU}{\mathcal{U}}
\newcommand{\myr}{q}
\newcommand{\Urho}{U_{\rho}}
\newcommand{\myalpha}{\alpha_*}
\newcommand{\bd}[1]{\mathbf{#1}}
\newcommand{\Eq}[1]{Eq.~(\ref{#1})}
\newcommand{\Eqn}[1]{Eq.~(\ref{#1})}
\newcommand{\Eqns}[1]{Eqns.~(\ref{#1})}
\newcommand{\Figref}[1]{Fig.~\ref{#1}}
\newtheorem{theorem}{Theorem}
\newcommand{\me}{\textrm{m}_{\textrm{e}}}
\newcommand{\sgn}{\textrm{sign}}
\newcommand*{\bfrac}[2]{\genfrac{\lbrace}{\rbrace}{0pt}{}{#1}{#2}}

\newcommand{\CTensorName}{transport shape-averaged }

\title{Space and Space-Time geodesics in Schwarzschild geometry}

\author{Lorenzo Resca}
\email{resca@cua.edu}
\homepage{http://physics.cua.edu/people/faculty/homepage.cfm}
\thanks{Corresponding Author.}

\affiliation{Department of Physics, 
The Catholic University of America,
Washington, DC 20064}

\date{\today}

\begin{abstract}
Geodesic orbit equations in the Schwarzschild geometry of general relativity reduce to ordinary conic sections of Newtonian mechanics and gravity for material particles in the non-relativistic limit. On the contrary, geodesic orbit equations for a proper spatial submanifold of Schwarzschild metric at any given coordinate-time correspond to an unphysical gravitational \textit{repulsion} in the non-relativistic limit. This demonstrates at a basic level the centrality and critical role of \textit{relativistic time} and its intimate pseudo-Riemannian connection with space. Correspondingly, a commonly popularized depiction of geodesic orbits of planets as resulting from the curvature of space produced by the sun, represented as a rubber sheet dipped in the middle by the weighing of that massive body, is mistaken and misleading for the essence of relativity, even in the non-relativistic limit.
\end{abstract}



\maketitle

\section{Introduction}  

A central element in Einstein's theory of general relativity (GR) is that ideal point-like test-particles follow time-like or null geodesics in a four-dimensional (4D) relativistic space-time which is curved as a pseudo-Riemannian manifold in connection with a stress-energy tensor derived from matter and radiation. A cosmological constant may further contribute to curvature in the corresponding Einstein field equations. We shall not be concerned with that constant or those equations in this paper, nor with any grand perspective that GR poses for our understanding of the cosmos over an immense range of scales. Far more modestly, I wish to return to basic concepts and remind students who already possess an introductory, but nonetheless physically and mathematically precise, understanding of GR that a commonly popularized depiction of geodesics in curved space alone is factually incorrect and conceptually misleading for the essence of relativity, even in the non-relativistic limit.

A most simplistic version of such misleading depiction suggests that bound orbits of planets around our sun are actually geodesics, i.e., non-turning intrinsically, or as `straight' as they can possibly be, in a surrounding space that is stretched and dipped like a rubber sheet by the massive weight of the sun in the middle. It is easy to demonstrate how profoundly mistaken that depiction really is, provided that the reader is proficient with GR concepts and techniques at the minimum level of introductory books, such as one of Schutz.\cite{Schutz2Ed} Thus I will show that using notations and equations of Ref.~\onlinecite{Schutz2Ed} without repeating their definitions and derivations in detail. 

Except for a few minor changes, this paper has just been published in Ref.~\onlinecite{Resca}.

\section{Schwarzschild geodesics and their non-relativistic limit}

The Schwarzschild coordinates,  
$x^{\mu} = (ct, r, \theta,\phi)$, also labeled with $\mu = 0, 1, 2, 3$, and metric line element

\begin{align}\label{metric}
ds^2  = & g_{\mu \nu} dx^{\mu} dx^{\nu} \nonumber\\
      = & -\bigg( 1- \frac{G}{c^2} \frac{2M}{r} \bigg)(c dt)^2 + \bigg(1 - \frac{G}{c^2} \frac{2M}{r} \bigg)^{-1} (dr)^2 + \nonumber\\
        & r^2 (d \theta)^2 + r^2 \sin^2 \theta (d \phi)^2,
\end{align}
are derived and discussed by Schutz, arriving at Eq. (10.36) in Ref.~\onlinecite{Schutz2Ed}. The one and only central parameter that characterizes GR curvature and introduces gravity in Schwarzschild metric, \Eq{metric}, is Schwarzschild radius, 

\begin{align}\label{radius}
r_{S} \equiv \frac{G}{c^2} 2 M.
\end{align}
Most interesting hystorical and technical information celebrating the centenary of Schwarzschild's two ground-breaking papers published in 1916 is provided in Ref.~\onlinecite{Schwarzschild(2016)}, for example. 

In order to simplify or shorten GR equations or relations, it is often convenient to express them in geometrized units by setting $c=1$ for the speed of light and $G=1$ for Newton's gravitational constant. For sake of greater transparency, we may retain general units in this paper, but unit conversions and comparisons can be easily performed by following the procedure outlined in Table 8.1 and in the corresponding subsection of Ref.~\onlinecite{Schutz2Ed}.

Schwarzschild metric describes space-time in the vacuum outside a spherical non-rotating star or black-hole singularity of mass $M$ at the origin. That metric is static, meaning that all metric tensor components, 
$g_{\mu \nu}$, are independent of the coordinate-time, $t$, and the geometry remains unchanged by time-reversal, $t \rightarrow -t$.

For a material point-like test-particle of vanishing mass $m>0$, we thus introduce the four-momentum $p^\mu = m \frac{d x^{\mu}}{d \tau}$, whose pseudo-norm
\begin{equation}\label{pseudo-norm}
  p_\mu p^\mu=g_{\mu \nu} p^\mu p^\nu =-m^2c^2
\end{equation}
defines the invariant proper-time interval $c{d \tau} = \sqrt{-ds^2}$. Such ${d \tau}>0$ represents the ticking of an ideal clock attached to the material test-particle.

The geodesic equation for momentum covariant components,
\begin{equation}\label{covariantgeodesic}
  m \frac{d p_{\beta}}{d \tau}
    = \frac{1}{2} \bigg( \frac{\partial g_{\nu \alpha}}{\partial x^{\beta}}  \bigg) p^\nu p^{\alpha},
\end{equation}
is derived by Schutz quite generally, arriving to Eq. (7.29) in Ref.~\onlinecite{Schutz2Ed}. For the static Schwarzschild metric, the time-component $(\beta=0)$ immediately provides conservation of energy,
\begin{equation}\label{conservationofenergy}
p_0 \equiv -E /c \equiv -m c \tilde{E}= g_{00} p^0 = - \bigg( 1-\frac{r_S} {r} \bigg) m c \frac{dt}{d \tau} =\mathrm{const}.
\end{equation}
The spherical symmetry of the Schwarzschild metric leads to planar geodesic orbits, which we may thus assume to be equatorial, maintaining constant the polar angle $\theta = \frac{\pi}{2} = \mathrm{const}$. Spherical symmetry and the geodesic \Eq{covariantgeodesic} for the azimuthal angle $\phi$, or $\beta=3$ component, implies further conservation of angular momentum value as 
\begin{equation}\label{angularmomentum}
  p_{\phi} \equiv L \equiv m \tilde{L}= g_{\phi \phi} p^{\phi}= r^2 m \frac{d \phi}{d \tau}= \mathrm{const}.
\end{equation}

Introducing these conserved quantities into the conservation of four-momentum pseudo-norm, \Eq{pseudo-norm}, we arrive at the geodesic equation of motion for radial distance,
\begin{equation}\label{geodesicradial}
   \bigg(\frac{d r}{d\tau}\bigg)^2
    = c^2 \tilde{E}^2 - c^2 +G\frac{2M}{r} - \frac{\tilde{L}^2}{r^2} + \frac{G}{c^2} \frac{2M}{r}   \frac{\tilde{L}^2}{r^2} ,
\end{equation}
as demonstrated by Schutz in Eq. (11.11) and illustrated in Fig. 11.1 of Ref.~\onlinecite{Schutz2Ed}.

We may further derive a geodesic \textit{orbit} equation in terms of the azimuthal angle, $\phi$, rather than the proper time, $\tau$, by dividing both sides of \Eq{geodesicradial} by $({d \phi}/{d \tau})^2$, as provided in \Eq{angularmomentum}. That yields
\begin{equation}\label{geodesicorbitradial}
  \bigg(\frac{d r}{d \phi}\bigg)^2
    = \frac{r^4}{\tilde{L}^2} \bigg\{c^2\tilde{E}^2 - c^2 +G\frac{2M}{r} - \frac{\tilde{L}^2}{r^2} + \frac{G}{c^2} \frac{2M}{r}   \frac{\tilde{L}^2}{r^2}  \bigg\}.
\end{equation}
The last term in \Eq{geodesicorbitradial} is responsible for the famous GR correction to the perihelion precession of Mercury's orbit: cf. Ref.~\onlinecite{Schutz2Ed}, pp. 287-291.

The non-relativistic limit of the GR geodesic orbit equation is obtained by assuming that $r>>r_S$ and that $r>>\left|\tilde{L}\right| / c$ for non-relativistic velocities. Thus we may correspondingly drop the last term in \Eq{geodesicorbitradial}, yielding

\begin{equation}\label{NRgeodesicorbitradial}
 \bigg(\frac{d r}{d \phi}\bigg)^2
    \simeq \frac{r^4}{\tilde{L}^2} \bigg\{c^2\tilde{E}^2 - c^2 +G\frac{2M}{r} - \frac{\tilde{L}^2}{r^2} \bigg\}.
\end{equation}

In the non-relativistic Newtonian limit we must also consider low energies relative to $m c^2$, implying that $|\tilde{E}^2-1| << 1$. We may thus rescale energy as 
\begin{equation}\label{relabeledenergy}
  \mathcal{E} \equiv \frac{1}{2} m (\tilde{E}^2-1)c^2,
\end{equation}
and assume that $|\mathcal{E}| << mc^2$ in the non-relativistic limit. We then arrive to the gravitational Newtonian orbit equation
\begin{equation}\label{Newtonianorbit}
  \bigg(\frac{d r}{d \phi} \bigg)^2
    \simeq \frac{ 2 m r^4}{L^2} \bigg\{\mathcal{E} + G \frac{M m}{r} - \frac{L^2}{2 m r^2}\bigg\} .
\end{equation}
This coincides with Eq. (3.12) in Ref.~\onlinecite{Fetter-Walecka}, for example. The Newtonian orbit \Eq{Newtonianorbit} is readily integrated to yield $r=r(\phi)$, which represents (arcs of) ellipses for $\mathcal{E}<0$, parabolae for $\mathcal{E}=0$, and one branch of hyperbolae for $\mathcal{E}>0$, having the center of force at $r=0$ located at the focus \textit{inside} that branch: see Fig. 5.2 in Ref.~\onlinecite{Fetter-Walecka}, for example.

\section{Geodesic orbit equations in space submanifolds of Schwarzschild metric}

A most natural way to consider proper space by itself in the Schwarzschild metric is to regard it as a three-dimensional (3D) submanifold at any given coordinate-time, $t$, exploiting the fact that the full space-time metric is static. This leads to a submetric line element for $x^{i} = (r, \theta, \phi)$ spatial coordinates, also labeled with $i = 1, 2, 3$, given by
\begin{align}\label{spatialmetric}
  d S^2 &= g_{ij} dx^i dx^j \nonumber\\
        &= \bigg(1 - \frac{r_S}{r} \bigg)^{-1} (dr)^2 + r^2 (d \theta)^2 + r^2 \sin^2 \theta (d \phi)^2.  
\end{align}
Since spatial coordinates are in fact space-like only outside the Schwarzschild radius, we are only interested here in that region within our event horizon, having $r>r_S$. In that region, $d S^2 > 0$ represents the line element of a 3D positive-definite Riemannian submetric.

We can parametrize curves in that 3D spatial submanifold with an affine parameter, $\lambda$, such that tangent vectors 
$V^i=\frac{d x^i}{d \lambda}$ have a positive-definite norm 
\begin{equation}\label{spatialnorm}
  V_i V^i= g_{ij} V^i V^j = C^2 > 0 .
\end{equation}
Geodesic curves in the 3D spatial submanifold then obey the equation
\begin{equation}\label{spatialgeodesic}
  \frac{d V_k}{d \lambda}= \frac{1}{2} \bigg( \frac{\partial g_{ij}}{\partial x^k} \bigg) V^i V^j .
\end{equation}

Spherical symmetry leads again to planar geodesic curves, which can be thus assumed to be equatorial, having $\theta = \frac{\pi}{2} = \mathrm{const}$. Spherical symmetry further leads from \Eq{spatialgeodesic} for $k=3$ to a conservation law of the form

\begin{equation}\label{spatialangularmomentum}
  V_{\phi} \equiv L =g_{\phi \phi} V^{\phi} = r^2 \frac{d \phi}{d \lambda} = \mathrm{const} .
\end{equation}
We should now consider whether to attribute to the $L$ constant in \Eq{spatialangularmomentum} some angular momentum interpretation, as we properly did in \Eq{angularmomentum}. Angular momentum depends on velocity, which depends on time. In our spatial submanifold, however, coordinate time is fixed, and geodesics can be parameterized only in terms of an affine parameter, $\lambda$, ultimately proportional to their arc lengths. Nevertheless, we can always reparameterize $\lambda$ by multiplication with any arbitrary constant, including one that gives to $\lambda$ dimensions of time over mass, thus making \Eq{spatialangularmomentum} at least dimensionally consistent with \Eq{angularmomentum}. We may further regard angular momentum as a generator of spatial rotations, which coincide in both the space-time metric and its spatial sub-metric, since both preserve the same spherical symmetry. From these perspectives, it is permissible to maintain a nomenclature or interpretation of the $L$ constant in \Eq{spatialangularmomentum} as at least proportional to some angular momentum value, which may be positive, negative, or zero. In fact, we are about to derive a spatial geodesic \textit{orbit} \Eq{spatialradialgeodesicorbit} that will eliminate any explicit appearance of the affine parameter $\lambda$, just as we did for the space-time geodesic \textit{orbit} \Eq{geodesicorbitradial}. Ultimately, we will be able to obtain a spatial geodesic \textit{orbit} \Eq{spatialradialgeodesicorbitperiastron} that depends on only two geometrical parameters, the orbit periastron, $r_{p}$, and the Schwarzschild radius, $r_{S}$, thus eliminating any need or appearance of angular momentum, $L$, and energy, $\mathcal{E}_s$, independently.

Combining \Eq{spatialangularmomentum} with conservation of norm, \Eq{spatialnorm}, we obtain
\begin{equation}\label{spatialgeodesicnorm}
  V_i V^i = V_r V^r + V_{\phi} V^{\phi} = \bigg( 1- \frac{r_S}{r} \bigg)^{-1}  \bigg( \frac{dr}{d \lambda}  \bigg)^2 + \frac{L^2}{r^2} =C^2 .
\end{equation}
This provides a geodesic equation for the radial coordinate in a spatial submanifold at any given time, $t$, namely
\begin{equation}\label{spatialradialgeodesic}
  \bigg(\frac{dr}{d\lambda}\bigg)^2
    = C^2- C^2 \frac{r_S}{r} -  \frac{L^2}{r^2} + \frac{r_S}{r}   \frac{L^2}{r^2}.
\end{equation}

Dividing both sides of \Eq{spatialradialgeodesic} by 
$\bigg(\frac{d \phi}{d \lambda} \bigg)^2$, as provided in \Eq{spatialangularmomentum}, we eliminate any explicit appearance of affine parameter and obtain once again a geodesic \textit{orbit} equation, expressed in terms of the azimuthal angle, $\phi$. Since $\frac{L^2}{C^2}$ has square-length units in \Eq{spatialradialgeodesic}, we have consistently
\begin{equation}\label{spatialradialgeodesicorbit}
\bigg( \frac{dr}{d \phi}  \bigg)^2= \frac{r^4}{L^2} \bigg\{ C^2- C^2 \frac{G}{c^2} \frac{2M}{r} -  \frac{L^2}{r^2} + \frac{G}{c^2}  \frac{2M}{r}   \frac{L^2}{r^2} \bigg\} .
\end{equation}

We may consider a weak-field limit of the GR geodesic orbit \Eq{spatialradialgeodesicorbit} in the spatial submanifold by assuming asymptotically large distances. Thus, ignoring the last term in \Eq{spatialradialgeodesicorbit}, we obtain 
\begin{equation}\label{NRspatialradialgeodesicorbit}
\bigg( \frac{dr}{d \phi}  \bigg)^2 \simeq \frac{r^4}{L^2} \bigg\{ C^2- C^2 \frac{G}{c^2} \frac{2M}{r} -  \frac{L^2}{r^2} \bigg\}.
\end{equation}

If we try to compare that orbit \Eq{NRspatialradialgeodesicorbit} with the Newtonian orbit \Eq{Newtonianorbit}, we have to associate the positive norm constant $C^2$ with $2m \mathcal{E}_s>0$, yielding
\begin{equation}\label{relabeledNRspatialradialgeodesicorbit}
  \bigg(\frac{dr}{d\phi}\bigg)^2
    \simeq \frac{ 2m r^4}{L^2} 
           \bigg\{\mathcal{E}_s - \bigg(\frac{2\mathcal{E}_s}{m c^2}\bigg)G\frac{Mm}{r}
                  -  \frac{L^2}{2m r^2} 
           \bigg\}. 
\end{equation}
By comparing signs relative to that of the familiar effective centrifugal repulsive potential, we may then conclude that the GR curvature of a Schwarzschild spatial submanifold at any given time, $t$, corresponds to a weakly \textit{repulsive} gravitational potential,
\begin{equation}\label{repulsivepotential}
  +\bigg( \frac{2  \mathcal{E}_s }{m c^2} \bigg) G \frac{Mm}{r}>0 ,
\end{equation}
in a weak-field limit. Corresponding orbits can thus only be arcs of hyperbolae. Furthermore, assuming a vanishing prefactor, $0<\frac{2  \mathcal{E}_s }{m c^2}<<1$, in the non-relativistic limit of the \textit{repulsive} gravitational potential, those orbits become virtually Euclidean straight lines, as expected for an essentially flat spatial submanifold of Schwarzschild metric in the non-relativistic limit. That is incompatible, for example, with bound orbits $(\mathcal{E}<0)$ for the \textit{attractive} and much stronger gravitational potential, 
\begin{equation}\label{attractivepotential}
  -G \frac{Mm}{r}<0 ,
\end{equation}
of the physically correct Newtonian mechanics in the non-relativistic limit of space-time GR, i.e., \Eq{Newtonianorbit}.

In any case, regardless of whether one may or may not wish to consider weak-field and/or non-relativistic limits, it is clear from direct comparison of the \textit{exact space-time geodesic orbit} \Eq{geodesicorbitradial} and the \textit{exact spatial-submanifold geodesic orbit} \Eq{spatialradialgeodesicorbit} that the former equation demands \textit{gravitational attraction} exclusively, whereas the latter equation invariably contains one term, namely its second, which corresponds to \textit{gravitational repulsion}!

\section{Null geodesics in space-time}

The pseudo-Riemannian Schwarzschild metric in space-time also admits \textit{null geodesics}, having $ds^2=0$ in \Eq{metric}. Null geodesics are travelled exclusively by exactly massless $(m \equiv 0)$ point-like test-particles, having a four-momentum $p^\mu = \frac{d x^{\mu}}{d \lambda}$ with null pseudo-norm
\begin{equation}
  p_\mu p^\mu= g_{\mu \nu} p^\mu p^\nu=0.
\end{equation}

Conservation of energy and angular momentum in their geodesic equation lead to a radial component equation
\begin{equation}\label{nullgeodesicradial}
  \bigg(\frac{dr}{d\lambda}\bigg)^2
    = \frac{E^2}{c^2} -\frac{L^2}{r^2} + \frac{G}{c^2} \frac{2M}{r} \frac{L^2}{r^2}.
\end{equation}
The spherical symmetry of the Schwarzschild metric has led again to planar equatorial geodesics, maintaining $\theta = \frac{\pi}{2} = \mathrm{const}$. In general units, \Eq{nullgeodesicradial} coincides with Eq. (11.12) in Ref.~\onlinecite{Schutz2Ed}. The corresponding geodesic \textit{orbit} equation is
\begin{equation}\label{nullgeodesicorbitradial}
  \bigg(\frac{dr}{d\phi}\bigg)^2 
    = \frac{r^4}{L^2}
      \bigg\{\frac{E^2}{c^2} -\frac{L^2}{r^2} + \frac{G}{c^2} \frac{2M}{r} \frac{L^2}{r^2} \bigg\}.
\end{equation}
The last term in \Eq{nullgeodesicorbitradial} is responsible for the GR deflection of star-light grazing the sun, first famously confirmed by observations in the eclipse of 1919: cf. Ref.~\onlinecite{Schutz2Ed}, pp. 293-295.

Estimating qualitatively the linear momentum of the massless test-particle, or `photon,' as $p^{r} = \frac{d r}{d \lambda}  \sim  \frac{E}{c}$, and its angular momentum as $L \sim r \frac{E}{c}$, we have $\frac{L^2}{r^2} \sim \frac{E^2}{c^2}$. For $r>> \frac{G}{c^2} M$, the last term in \Eq{nullgeodesicradial} or in \Eq{nullgeodesicorbitradial} becomes asymptotically smaller than each of its two preceding terms. One may neglect that last term in the weak-field limit, which thus reduces to flat space-time. Thus, for $r>> \frac{G}{c^2} M$, the massless $(m \equiv 0)$ photon is no longer subject to any gravitational potential. The photon essentially moves along a straight Euclidean line as in flat space-time. This contrasts with the much older prediction of some star-light deflection by grazing the sun made by Cavendish (1784) and Soldner (1801) based on a purely Newtonian description of light particles: cf. Ref.~\onlinecite{Berry}, Sec. 5.4, pp. 85-88, and Ref.~\onlinecite{Will}. 

However one may regard or disregard qualitative weak-field estimates, it is evident that the exact geodesic orbit \Eq{nullgeodesicorbitradial} for the massless $(m \equiv 0)$ photon is precisely missing the attractive Newtonian gravitational potential that is instead prominent in the exact geodesic orbit \Eq{geodesicorbitradial} for a material point-like test-particle of mass $m>0$, independently of how small or `vanishing' that mass may be.

If we consider instead the spatial submanifold of Schwarzschild metric at any given coordinate-time, $t$, its line element $d S^2 > 0$, as given in \Eq{spatialmetric}, corresponds to a positive-definite Riemannian metric within the event horizon, where $r>r_S$. Therein, null geodesics cannot exist on the spatial submanifold, by definition.

%
%
%
%

\section{Flamm's paraboloid}

There is in fact a rigorous procedure to depict the metric of space alone as a submanifold of Schwarzschild metric at any given time, $t$, which can be summarized as follows. First of all, let us consider only a two-dimensional (2D) space to represent equatorial planes, maintaining $\theta = \frac{\pi}{2} = \mathrm{const}$. The corresponding line element in \Eq{spatialmetric} thus reduces to 
\begin{equation}\label{reducedspatialmetric}
  d S^2= \bigg(1 - \frac{r_{S}}{r} \bigg)^{-1} (dr)^2 + r^2 (d \phi)^2  
\end{equation} 
for the $(r,\phi)$ coordinates. Let us then embed the corresponding 2D submanifold in the ordinary 3D Euclidean space, by associating $r^2$ with $(X^2 + Y^2)$ and by defining, for $r \ge r_{S}$,
\begin{equation}\label{Flamm}
  Z^2= 4 r_{S}^2\bigg(\frac{r}{r_{S}} - 1 \bigg).  
\end{equation} 
This is known as Flamm's paraboloid of revolution about the $Z-$axis.\cite{Flamm} It derives from straightforward integration after setting $d S^2= (d Z)^2 + (d r)^2 + r^2 (d \phi)^2$ equal to $d S^2$ in \Eq{reducedspatialmetric}. Pictures and discussions related to Flamm's paraboloid are ubiquitously provided, including classic textbooks such as: Ref.~\onlinecite{MTW}, p. 837, Fig. 31.5; Ref.~\onlinecite{Rindler}, pp. 136-142; Ref.~\onlinecite{Wald}, p. 155, Fig. 6.10; Ref.~\onlinecite{Berry}, pp. 70-73, Fig. 37; Ref.~\onlinecite{Schutz2Ed}, p. 257, Fig. 10.1, even sketched on the cover of that book's first Edition. Ironically, it is precisely a misguided interpretation of such correct pictures that is mainly responsible for the confusion of popular rubber-sheet analogies of curved space. 

For $L = 0$, it is easy to show that the geodesic \Eq{spatialradialgeodesic} leads to corresponding parabolic curves as given in \Eq{Flamm}. These are geodesic orbits for which the test-particle is radially directed, thus maintaining $\phi = \mathrm{const}$.

For $L \ne 0$, it is possible to prove that the geodesic orbit \Eq{spatialradialgeodesicorbit} admits \textit{no circular orbit solutions} with $r = r_0 = \mathrm{const} > r_{S}$. In fact, \textit{the only bound solution is the minimal circle with} $r = r_{S}$, which is the unstable geodesic orbit that encircles the `throat' of Flamm's paraboloid at $Z=0$. Other than that, all other circles commonly drawn at various $r= \mathrm{const} > r_{S}$ do \textit{not} represent geodesic orbits on Flamm's 2D spatial submanifold for any arc length. The geometry of Flamm's paraboloid is in fact hyperbolic-like, with a negative intrinsic Gaussian curvature $K = - \frac{G}{c^2} \frac{M}{r^3}$ that quickly vanishes for $r>>r_{S}$, rapidly reaching the asymptotic limit of flat 2D space.\cite{Rindler} Therein, geodesic orbits become virtually Euclidean straight lines, while still asymptotically bending \textit{away} from the `throat' of Flamm's paraboloid.

The geodesic orbit \Eq{spatialradialgeodesicorbit} admits only a single turning point, obtained by equating \Eq{spatialradialgeodesicorbit} to its minimum zero value. One can then express the orbit periastron as 
\begin{equation}\label{periastron}
r_{p}^2 = {\frac{L^2}{C^2}} \equiv \frac{L^2}{2 m \mathcal{E}_s} ,
\end{equation}
for any $r_{p} > r_{S}$. One may then re-express the orbit \Eq{spatialradialgeodesicorbit} exclusively in terms of $r_{p}$ and $r_{S}$ as 

\begin{equation}\label{spatialradialgeodesicorbitperiastron}
\bigg( \frac{dr}{d \phi}  \bigg)^2= \frac{r^4}{r_{p}^2} -  \frac{r_{S} r^3}{r_{p}^2} - r^2 + r_{S} r.
\end{equation} 

The expression of the geodesic orbit in \Eq{spatialradialgeodesicorbitperiastron} thus depends on a single initial condition, specified by the periastron, $r_{p}$. That originates from the fact that we can reparameterize the affine parameter $\lambda$ in \Eq{spatialangularmomentum} by multiplication with an arbitrary constant. Thus, only the ratio $\frac{L^2}{C^2}$ in $r_{p}^2$ ultimately matters, rather than the $L^2$ and $C^2$ constants independently. This is already implicit in \Eq{spatialradialgeodesicorbit}. Still, it is typically more instructive or convenient to study or solve spatial geodesic orbit equations in the form of \Eq{spatialradialgeodesicorbitperiastron}, whether analytically or numerically. 

It is possible to integrate the orbit \Eq{spatialradialgeodesicorbitperiastron} for $L \ne 0$ following a standard procedure, involving a separation of variables, followed by a functional inversion, which yields $r = r(\phi)$: see, for instance, p. 13 in Ref.~\onlinecite{Fetter-Walecka}. If we drop the last and most relativistic term in \Eq{spatialradialgeodesicorbitperiastron}, we can directly derive the analytic solution, representing a branch of hyperbola. The convex angle comprised by its asymptotes varies from a maximum of $\pi$, corresponding to a straight line, to a minimum of 126.87 degrees. 

Relativistically however, if $r_{p}$ becomes of the order of $r_{S}$, while still maintaining $r_{p} > r_{S}$, spatial geodesic orbits become much more complicated, especially at short distances. In that situation, the second and fourth terms in \Eq{spatialradialgeodesicorbitperiastron}, representing gravitational repulsion and attraction, respectively, produce comparable and competing effects. It is always possible, however, to obtain an analytic solution in terms of elliptic integrals. In fact, we have generated and studied many such orbits numerically and analytically. Detailed results and discussions are beyond the scope of this paper and will be reported elsewhere.\cite{Rafael} Remarkably, whether encircling the `throat' of Flamm's paraboloid once or multiple times or not at all, the angle comprised between the orbit asymptotes turns out to be always \textit{concave}, varying from a minimum of $\pi$, corresponding to a straight line in the non-relativistic limit, to a maximum of 2$\pi$ when $r_{p}$ is appropriately close to $r_{S}$. 

\section{Interpretation of results}

We wish of course to understand more precisely the physical and mathematical origins of the discrepancy between GR geodesic orbits in space-time and those in space alone, which strikingly persists even in the non-relativistic limit of GR: cf. \Eq{Newtonianorbit} and \Eq{relabeledNRspatialradialgeodesicorbit}, \Eq{repulsivepotential} and \Eq{attractivepotential}. Clearly, the central element is that geodesics of material test-particles are \textit{time-like} in GR space-time, as shown by the \textit{negative pseudo-norm} of their tangent vectors, \Eq{pseudo-norm}, whereas geodesics in the spatial submanifold of Schwarzschild metric at any given coordinate-time, $t$, are \textit{space-like}, as shown by the \textit{positive norm} of their tangent vectors, \Eq{spatialnorm}, for $r>r_S$.

From another perspective, conservation of energy, \Eq{conservationofenergy}, is associated with invariance under time-translations, which strictly applies only to geodesics in Schwarzschild space-time geometry. A proper spatial submanifold of that can be most sensibly considered by assuming \textit{simultaneity}, i.e., $d t = 0$, in Schwarzschild's static metric. However, this simultaneity condition excludes any bona fide conservation of energy for geodesics constrained to that spatial submanifold. Thus, the \textit{positive norm}, $C^2$, of tangent vectors to space-constrained geodesics can only be formally associated with a fictitious `space-invariant' energy, $2m \mathcal{E}_s>0$, resulting in the non-relativistic limit of \Eq{relabeledNRspatialradialgeodesicorbit}. Only that allows to interpret the spatial geodesic orbit as a gravitational orbit, but with the critical difference of having a \textit{weakly repulsive potential}, as given in \Eq{repulsivepotential}.

From yet a third perspective, notice that the derivation of geodesic equations in space-time, \Eq{geodesicradial} and \Eq{geodesicorbitradial}, even in the non-relativistic limit, \Eq{NRgeodesicorbitradial}, requires consideration of both \textit{time-like}, $g_{tt}$, and \textit{space-like}, $g_{rr}$, metric tensor components on an \textit{equal footing}. On the contrary, the derivation of geodesic equations in the spatial submanifold, \Eq{spatialradialgeodesic}, \Eq{spatialradialgeodesicorbit}, and \Eq{NRspatialradialgeodesicorbit}, completely excludes consideration of $g_{tt}$, having required $d t=0$ at the outset.  

All these related perspectives indicate that what is critically missing from the space-only descrition of GR curvature is the fundamental concept of \textit{relativity of time and simultaneity, and its pseudo-Riemannian connection to the relativity of space}. Of course, the Newtonian account of gravity disregards absolutely that very concept. According to Newton, time and simultaneity are presumed to be absolute, while gravity is supposed to act instantly at all distances. Of course we currently know that it takes minutes or hours for the sun to influence gravitationally its planets while they move closer or further around it. Thus, in retrospect of course, one might wonder why Newtonian mechanics and gravity, absolutely defying such fundamental relativistic principle of space-time connection, could have worked so well from the beginning.

Our derivations and equations may help to figure that out more precisely. First of all, by deriving \textit{orbit} equations, we have avoided the relativity of time evolution to appear explicitly. Secondly, by considering weak-gravity regions, for example in the solar sytem, where $r>> \frac{G}{c^2} M \sim 1.5$\,km, we have limited ourselves to a nearly flat space-time. In the geodesic orbit \Eq{Newtonianorbit} for that space-time manifold, all three terms in curly brackets are comparably small and of the order of $|\mathcal{E}|<<m c^2$ along typically bound planetary orbits. By comparison to their planetary velocities, the speed of light is practically infinite. We may thus better realize why GR principles of space-time relativity and curvature are not altogether incompatible with Newtonian absolute principles in the appropriate non-relativitic limit.

On the other hand, why the $d t = 0$ denial of the relativity of time and assertion of simultaneity dooms from the outset a spatial submanifold consideration of the Schwarzschild metric as a possible explanation of Newtonian gravity in the non-relativistic limit? Evidently, the order in which certain limits are taken does matter. One thing is to derive a correct geodesic orbit equation in GR space-time, \Eq{geodesicorbitradial}, and then take its non-relativistic limit, \Eq{NRgeodesicorbitradial}. An altogether different matter is to consider a spatial submanifold of Schwarzschild metric which eliminates from the outset the fundamental principle of relativity of space-time, simultaneity, and curvature as their pseudo-Riemannian connection. In dealing with `nearly flat' space-time, as weak as curvature and gravity may be, limits must be taken in the appropriate order, beyond the zero-order of `absolutely flat' space-time. 

\section{Alternative approaches to spatial curvature}

Of course over time there have been many different accounts of spatial curvature, as it may relate to gravity separately from full space-time curvature. Reasonably simple approaches have been discussed in Refs.~\onlinecite{Price, Ellingson, Gruber}, for example. It may thus be useful to recast at least parts of such accounts in terms of the covariant geodesic orbit formulation, based on \Eq{covariantgeodesic}, that I have consistently developed throughout this paper.

Let us first consider an approximate 4D pseudo-Riemannian manifold with metric 

\begin{align}\label{metricwithoutspatialcurvature}
ds^2  = & g_{\mu \nu} dx^{\mu} dx^{\nu} \nonumber\\
      = & -\bigg( 1-\frac{r_S}{r} \bigg)(cdt)^2 + (dr)^2 + \nonumber\\
        & r^2 (d \theta)^2 + r^2 \sin^2 \theta (d \phi)^2.
\end{align}
This differs from the physically correct Schwarzschild metric in that the 3D spatial submanifold at any given coordinate-time, $t$, has been devoided of any curvature in \Eq{metricwithoutspatialcurvature}. Following the same procedures that I adopted earlier produces here the geodesic orbit equation: 

\begin{equation}\label{exactnospatialcurvatureradialgeodesicorbit}
  \bigg(\frac{d r}{d \phi}\bigg)^2
    = \frac{r^4}{\tilde{L}^2} \bigg\{c^2\tilde{E}^2 \bigg(1 - \frac{G}{c^2} \frac{2M}{r} \bigg)^{-1} - c^2  - \frac{\tilde{L}^2}{r^2}  \bigg\}.
\end{equation}
In the weak-field limit, $r>> \frac{G}{c^2} M$, the geodesic orbit \Eq{exactnospatialcurvatureradialgeodesicorbit} reduces to

\begin{equation}\label{nospatialcurvatureradialgeodesicorbit}
  \bigg(\frac{dr}{d\phi}\bigg)^2
   \simeq \frac{ 2m r^4}{L^2} 
           \bigg\{\mathcal{E} + G\frac{Mm}{r} + \bigg(\frac{2\mathcal{E}}{m c^2}\bigg)G\frac{Mm}{r}
                  -  \frac{L^2}{2m r^2} 
           \bigg\}, 
\end{equation}
where we have relabeled the energy according to \Eq{relabeledenergy}.

For $|\mathcal{E}|<<m c^2$, this geodesic orbit \Eq{nospatialcurvatureradialgeodesicorbit} coincides with Schwarzschild result in the non-relativistic Newtonian limit, \Eq{Newtonianorbit}, including the correct attractive gravitational potential, \Eq{attractivepotential}. This provided a major breakthrough from both physical and historical perspectives. It confirmed to Einstein that Newtonian gravity basically derives from the equivalence principle and its association with the gravitational redshift, even without full knowledge of Einstein field equations: cf. Chap. 18 of Ref. \onlinecite{SchutzGravity}, for example.
 
Let us consider alternatively a fictitious 4D pseudo-Riemannian manifold with metric 
 
 \begin{align}\label{metricwithouttimecurvature}
ds^2  = & g_{\mu \nu} dx^{\mu} dx^{\nu} \nonumber\\
      = & -(cdt)^2 + \bigg(1 - \frac{r_S}{r} \bigg)^{-1} (dr)^2 + \nonumber\\
        & r^2 (d \theta)^2 + r^2 \sin^2 \theta (d \phi)^2.
\end{align}
This differs from the Schwarzschild metric in that the time-like metric tensor component is assumed to be the same as it is in special relativity, i.e., $g_{tt} = -1$, whereas the 3D spatial submanifold at any given coordinate-time, $t$, maintains the same curvature as in Schwarzschild metric. Following the same procedures that we adopted earlier produces now the following geodesic orbit equation: 

\begin{equation}\label{notimecurvatureradialgeodesicorbit}
  \bigg(\frac{dr}{d\phi}\bigg)^2
    = \frac{ 2m r^4}{L^2} 
           \bigg\{\mathcal{E} - \bigg(\frac{2\mathcal{E}}{m c^2}\bigg)G\frac{Mm}{r}
                  -  \frac{L^2}{2m r^2} + \frac{G}{c^2}\frac{M}{r}\frac{L^2}{m r^2}
           \bigg\}. 
\end{equation}

Whether coincidentally or not, this geodesic orbit \Eq{notimecurvatureradialgeodesicorbit} formally coincides with \Eq{spatialradialgeodesicorbit} that I previously obtained for the spatial submanifold of Schwarzschild metric at any given coordinate-time, $t$. There is an important distinction, however. My previous geodesic orbit \Eq{spatialradialgeodesicorbit} had a \textit{space-like} origin. Thus I was bound to associate the positive constant $C^2$ with a positive energy term $2m \mathcal{E}_s>0$. The geodesic orbit \Eq{notimecurvatureradialgeodesicorbit} has a \textit{time-like} origin. Hence, its energy $\mathcal{E}$ may also be \textit{negative}, thus yielding a weakly \textit{attractive} gravitational potential. However, unbound orbits, allowing $r \rightarrow \infty $, require $\mathcal{E}>0$, which brings us back to the problem of a weakly \textit{repulsive} gravitational potential. Further analysis shows that this attractive/repulsive switching of the potential can occur only for $L=0$. For $L \ne 0$, by equating the orbit \Eq{notimecurvatureradialgeodesicorbit} to its minimum zero value, one can show that there is again only a single turning-point, which is a periastron that satisfies 
\begin{equation}\label{notimecurvatureperiastron}
r_{p}^2 = \frac{L^2}{2 m \mathcal{E}} 
\end{equation}
for any $r_{p} > r_{S}$. Then the corresponding energy $\mathcal{E}$ must again be \textit{positive}, yielding a weakly \textit{repulsive} gravitational potential corresponding to that of \Eq{repulsivepotential}.

It is also possible to determine \textit{null} geodesic orbits for the `splittable space-time' metric given in \Eq{metricwithouttimecurvature}. I obtain

\begin{equation}\label{nullgeodesicorbitradialsplittablespacetime}
  \bigg(\frac{dr}{d\phi}\bigg)^2 
    = \frac{r^4}{L^2}
      \bigg\{\frac{E^2}{c^2} -\bigg(\frac{E^2}{c^2}\bigg)\frac{G}{c^2} \frac{2M}{r} -\frac{L^2}{r^2} + \frac{G}{c^2} \frac{2M}{r} \frac{L^2}{r^2} \bigg\}.
\end{equation}

Whether coincidentally or not, also this \textit{null} geodesic orbit \Eq{nullgeodesicorbitradialsplittablespacetime} for the `splittable space-time' metric structurally coincides with my \textit{space-like} geodesic orbit \Eq{spatialradialgeodesicorbit} for the Schwarzschild spatial submanifold, if we let the positive constant $C^2$ in \Eq{spatialradialgeodesicorbit} correspond to the positive constant $\frac{E^2}{c^2}$ in \Eq{nullgeodesicorbitradialsplittablespacetime} in this case.

The \textit{null} geodesic orbit \Eq{nullgeodesicorbitradialsplittablespacetime} for `splittable space-time' critically differs from the \textit{exact null geodesic orbit} \Eq{nullgeodesicorbitradial} of Schwarzschild space-time metric, which rules out the second term within the curly brackets of \Eq{nullgeodesicorbitradialsplittablespacetime}. For light grazing the sun, our numerical integrations of the equivalent \Eq{spatialradialgeodesicorbit} and \Eq{nullgeodesicorbitradialsplittablespacetime} indicate a `spatial bending' of about half the total inward light deflection of 1.75 arc-seconds, which we recover for the exact null geodesic orbit \Eq{nullgeodesicorbitradial} of Schwarzschild space-time metric.\cite{Rafael} Our numerical integrations of null geodesics for the fictitious metric of \Eq{metricwithoutspatialcurvature} also indicate a `gravitational red-shift bending' of about half the total inward light deflection of 1.75 arc-seconds.\cite{Rafael} Other authors may have reached similar conclusions by different methods.\cite{Price, SchutzGravity} 

It may seem curious that both \textit{time-like} and \textit{null} geodesic orbits for the `splittable space-time' metric essentially coincide with \textit{space-like} geodesic orbits for the proper spatial Schwarzschild sub-metric, for example. In fact, it is possible to understand precisely all such matters by keeping track explicitly of all $g_{tt}$ and $g_{rr}$ factors and all norm, pseudo-norm or null terms in the exact derivation of geodesic orbits for all metrics considered. A critical feature is that the product of $g_{tt}$ and $g_{rr}$ is constant only for the exact Schwarzschild space-time metric, but not for the fictitious metrics of \Eq{metricwithoutspatialcurvature} and \Eq{metricwithouttimecurvature}. Having $g_{tt}g_{rr}=-1$, as in Minkowski's space-time, tells us that time and space bend \textit{inversely, relatively} to each other, in Schwarzschild space-time. That reflects a central requirement of the equivalence principle, namely, that the speed of light must remain a universal constant in any local freely-falling Lorentzian frame, in curved space-time, just as it is in flat space-time. 

\section{Some historical remarks and conclusions}

It was clear to Euclid, if not before, that the space around us may or may not be absolutely flat. Thus Euclid did not assert that he could mathematically demonstrate his fifth postulate on the basis of his other geometrical postulates or elements. In subsequent centuries, many mathematicians and natural philosophers tried hard to either prove mathematically or demonstrate practically that space could or could not be perfectly flat or `Euclidean.' Gauss's Theorema Egregium and his famous geodetic experiments with light rays led the way to mathematically rigorous theories of non-Euclidean geometries: cf. Ref.~\onlinecite{Berry}, p. 61, pp. 160-163. The formulation of Riemannian manifolds and geometry represents a crowning achievement and a momentous breakthrough in that quest. Remarkably, Riemann himself attempted to apply his geometry to configurational spaces in Lagrangian mechanics including the influence of external gravitational fields, but his efforts were doomed to failure in that regard: cf. Ref.~\onlinecite{Rindler}, Sec. 7.4, pp. 114-117. What Riemann did not know around 1854 was of course the theory of special relativity and the formulation of Minkowski's space-time in particular. Einstein figured all that and how to put it together to formulate a general relativity theory for a pseudo-Riemannian space-time that accounts for gravity as a manifestation of its curvature and connection.\cite{Schwarzschild(2016)} 

Now just about any intelligent person can at least in principle understand these ideas, among the grandest, if not the grandest, of all times. However, it still takes major training in both physics and mathematics to get to that point. Straightforward applications that I worked out in this article may have been evident to experts for more than a century. Still, this should remind us that `subtle is the Lord' and that one should `make a theory as simple as nature allows, but not simpler.' Accordingly, `perceptual visualization' of space alone as a rubber sheet deformed by the weight of the sun in the middle is deceiving and should not be used to suggest that planets merely follow geodesic orbits in a curved space. What is needed is a much deeper physical and mathematical appreciation of the relativistic connection between \textit{time} and \textit{space}, which is ultimately a consequence of the constancy of the speed of light in local freely-falling Lorentzian frames. 

A computer-generated depiction of the popular rubber sheet pinched and pulled down at its center has been featured in a NOVA program on `Black Hole Apocalypse,' broadcasted on January 10, 2018, on PBS.\cite{NOVA} A commentary describing that depiction includes the following excerpts: `According to Einstein, the apple and the Space Station and the astronauts are all falling freely along a curved path in space. And what makes that path curved? The mass of the earth ... So, according to Einstein, the mass of every object causes the space around it to curve ... All objects in motion follow the curves in space. So, how does the earth move the apple without touching it? The earth curves space and the apple falls freely along those curves. That, according to Einstein's general theory of relativity, is gravity: curved space. And that understanding of gravity, that an object causes the space around it to curve, leads directly to black holes.' 

Popular commentaries such as this may contradict not only Einstein's genius, but also common sense: see pp. 59-61 and Fig 26 in Ref.~\onlinecite{Berry}, for example. Once space is statically curved, any object that is released from any given point in any given direction should follow the same geodesic curve in such \textit{a-temporally curved space}. Likewise, a particle constrained to move on the surface of a sphere should follow a geodesic great circle, regardless of its encircling rate. Instead, we see that the object follows vastly different trajectories, depending critically on the speed with which the object is initially released. In fact, at the surface of the earth, the curvature of \textit{a-temporal space} is minuscule ($K \simeq - 1.7$ x 10$^{-27}$ cm$^{-2}$) and practically undetectable, as demonstrated by observations from Euclid to Gauss, and up to the most advanced current technologies. Indeed, at the surface of the earth, a particle approaching or traveling at the speed of light practically follows a Euclidean straight line. The vastly different curvatures in trajectories of objects or projectiles thrown at different speeds on or around the surface of the earth, which are part of our every-day experience, are consequence of \textit{space-time curvature}, not of \textit{a-temporal space curvature}. Except for this last but most important qualification, that central point is made repeatedly in the NOVA program on `Black Hole Apocalypse,' which is outstanding.\cite{NOVA}

In any event, recall that even for absolutely flat or \textit{Euclidean space} the fictitious 4D pseudo-Riemannian metric of \Eq{metricwithoutspatialcurvature} for \textit{curved time} correctly reduces to the non-relativistic Newtonian limit, thus correctly predicting all parabolic orbits that we ordinarily observe for projectiles thrown at non-relativistic speeds at the earth surface. On the contrary, the alternative fictitious 4D pseudo-Riemannian metric for exclusively \textit{curved space}, \Eq{metricwithouttimecurvature}, does \textit{not} predict such correct parabolic orbits in the non-relativistic limit.

Popular rubber-sheet funnel depictions of gravity and corresponding commentaries are often flawed from a rigorous scientific perspective. I showed that by reviewing how the Schwarzschild metric in relativistic \textit{space-time} yields geodesic orbit equations in space for $r=r(\phi)$ that reproduce all the correct conic sections for a Newtonian attractive gravitational potential, $-G \frac{Mm}{r}$, in the non-relativistic limit. On the contrary, considering the curvature of \textit{space} exclusively, obtained as a proper submanifold of the Schwarzschild metric at any given coordinate-time, $t$, yields geodesic orbit equations for $r=r(\phi)$ that correspond to a fictitious \textit{anti-gravitational} potential, $+\bigg( \frac{2  \mathcal{E}_s }{m c^2} \bigg) G \frac{Mm}{r}$, weakly \textit{repulsive} for material test-particles in the non-relativistic limit.

In the following Appendix, I briefly refer to a more advanced and general formulation of the differential geometry of Lorentzian manifolds that underlies explicit results obtained in this paper for the Schwarzschild geometry. I also briefly refer to further literature of broader historical and general interest.

\acknowledgments

I am grateful to Drs. Rafael T. Eufrasio, Nicholas A. Mecholsky, and Prof. Francesco Sorrentino for critical discussions, numerical results and technical support. I acknowledge critical contributions of anonymous reviewers of the European Journal of Physics that I have included in my paper and further summarized in the Appendix. 

\section{Appendix}

In the formalism of modern differential geometry of Lorentzian manifolds, i.e., manifolds that are equipped with a metric that is locally Minkowskian, static space-times, like Schwarzschild's, are warped products of a 3D Riemannian manifold as the base, modeling space, and the real line as the fiber, modeling time. It is a property of static space-times that geodesics, i.e., trajectories of freely falling test particles, do not typically correspond, i.e., they are not projected onto, geodesics of `fixed' space. See, for instance, Chapter 7, pp. 204-209, and Chapters 12-13, pp. 360-371, of Ref.~\onlinecite{O'Neill}. The main purpose of this paper has been to demonstrate this general result by relatively simple means, showing by explicit calculations for particular but important examples that orbits of test particles moving along space-time geodesics are different from orbits of points moving along geodesics in the `a-temporal' space of Schwarzschild's geometry.

Further relevant references of greater or deeper historical or technical interest include the following. Concerning Riemann, an English translation of his famous inaugural lecture can be found in Ref.~\onlinecite{Spivak}. Extensive historical accounts on the developments of space, time and gravitation theory, beginning with Gauss's, Riemann's and Clifford's ideas, are provided in Ref.~\onlinecite{Mehra}, pp. 92-178, and Ref.~\onlinecite{Jammer}, Chapter 5. Clifford's seminal contributions are included in Ref.~\onlinecite{Newman}, pp. 546-569, and more extensively in Ref.~\onlinecite{Clifford}.

Some conclusions similar to those of this paper have been reached recently from a different approach.\cite{Dadhich} In particular, it has been noted that geodesics with zero energy in 3D space coincide with null geodesics in 4D space-time. That is also the case in the \textit{limit} of vanishing energies in \Eq{spatialradialgeodesicorbit} and \Eq{nullgeodesicorbitradial}. 

\bibliographystyle{apsrev4-1}


\end{document}